# Collapse on the line – how synthetic dimensions influence nonlinear effects

André L. M. Muniz[1], Martin Wimmer[1], Arstan Bisianov[1], Roberto Morandotti[2,3,4] and Ulf Peschel[1]

1. Institute of Solid State Physics and Optics, Abbe Center of Photonics, Friedrich Schiller University Jena, Max-Wien-Platz 1, 07743 Jena, Germany
2. INRS EMT, 1650 Blvd Lionel Boulet, Varennes, PQ J3X 1S2, Canada
3. ITMO University, St. Petersburg, Russia
4. Institute of Fundamental and Frontier Sciences, University of Electronic Science and Technology of China, Chengdu 610054, China

**Introductory paragraph**

Power induced wave collapse is one of the most fascinating phenomena in optics as it provides extremely high intensities, thus stimulating a range of nonlinear processes. For low power levels, propagation of beams in bulk media is dominated by diffraction, while above a certain threshold self-focusing is steadily enhanced by the action of a positive nonlinearity. An autocatalytic blow-up occurs, which is only stopped by saturation of the nonlinearity, material damage or the inherent medium discreteness. In the latter case, this leads to energy localization on a single site. It is commonly believed that for cubic nonlinearities, this intriguing effect requires at least two transverse dimensions to occur and is thus out of reach in fiber optics. Following the concept of synthetic dimensions, we demonstrate that mixing short and long-range interaction resembles a two-dimensional mesh lattice and features wave collapse at mW-power levels in a genuine 1D system formed by coupled fiber loops.

**Manuscript**

Light-matter-interaction yields fascinating phenomena based on the interplay between linear wave spreading and nonlinearity[1], where the latter one critically depends on the dimensionality. Pulses propagating in optical fibers, a genuine one-dimensional (1D) system, tend to form solitons, an effect well described by the nonlinear Schrödinger equation (NLS)[2]. Here the balance of nonlinearity and dispersion imposes an unambiguous relationship between amplitude, energy and width of stable pulses. In contrast to fiber solitons, stationary states of the two-dimensional (2D) NLS can be scaled arbitrarily in amplitude and width, while energy stays constant[3,4]. This flexibility paves the ground for a catastrophic collapse in a finite time span[5]. In such situation, the beam compresses during propagation, while the amplitude increases, and consequently enhances the effect of the cubic nonlinearity in an autocatalytic way, ultimately ending in a blow-up of the nonlinear wave[2,3]. This process is stopped by material damage or saturation of the nonlinear action[6], thus potentially leading to stable solitons for quadratic[7], thermal[8] or photorefractive nonlinearities[9,10] (for an overview see[11]). Collapse also finds its end when the field structure is so far compressed that it resolves the discreteness of the underlying system, resulting in the formation of an extremely spiked single lattice site soliton[12,13,14,15]. In 2D waveguide arrays, discrete solitons[4,16] and even light bullets compensating also for temporal dispersion were reported[17].

Those discrete optical systems have recently become a valuable tool to simulate various lattice models and to investigate challenging and unexploited physical phenomena, such as PT-symmetry[18], exceptional points[19], unidirectional invisibility[20] or topological insulator lasers[21]. Whereas most of the proposals rely on the coupling of a limited number of local states to add a new degree of freedom[22,23], we here combine short and long range interaction following a scheme originally developed for quantum walks[24,25]. Our final aim is to investigate how synthetic dimensions influence nonlinear effects.

Following the Vlasov-Petrishchev-Talanov theorem[2], collapse only occurs if $(p-1)D \geq 4$, with $p$ being the order of the nonlinearity and $D$ the dimension of the system. In particular, collapse and the resulting boost of nonlinear phenomena is not expected for genuine one-dimensional fiber systems ($D=1$) with cubic nonlinearity ($p=3$). Here, following the introduced concept of synthetic dimensions, we experimentally demonstrate that a genuine 1D system formed by coupled fiber loops can resemble an extended effective lattice with two artificial transverse dimensions and a suitable band structure to enable the investigation of nonlinear propagation effects, leading to a collapse wave.

Our experimental set-up (see Fig. 1a) is entirely built from standard telecommunication equipment operated at a wavelength of 1550 nm. Two pairs of slightly dissimilar loops, each containing roughly 30 km of optical fiber, are coupled in order to create a synthetic mesh lattice encoded in the pulse arrival time (see Supplementary Note 2 and 3). An initial pulse is inserted via a fiber optical coupler in the outer left loop and splits into two pulses in the first 50/50 coupler at the entrance of the two inner loops (step I). The length of the inner two loops $L_1$ and $L_2$ differs slightly by $\Delta L_{\text{inner}} = 600$ m. After passing the inner pair of loops, the two pulses therefore arrive with a mutual delay of approximately $\Delta T_{\text{inner}} = 3$ µs at a second 50/50 coupler. Again, they split into now four pulses (see Fig. 1b), which propagate throughout the outer loops having a much shorter length difference of $\Delta L_{\text{outer}} = 6$ m, corresponding to a mutual delay time of $\Delta T_{\text{outer}} = 30$ ns (step II). After a mean round trip time $\bar{T} \approx 300$ µs, an ordered sequence of pulses returns to the first coupler and starts its journey again (see Fig. 1c). For any $m$ round trips, the pulses arrive in the photodetectors at different arrival times as

$$T_{\text{arrival}}(m, x, y) = m\overline{T} + x\, \Delta T_{\text{inner}} + y\, \Delta T_{\text{outer}}. \tag{1}$$

where $x$ and $y$ count how often the pulse has passed the longer than the shorter inner and outer loops, respectively. This condition allows for a unique identification of $x$ and $y$ for each pulse based on its arrival time provided that the length differences of the loops are sufficiently distinct and that the number of time steps $m$ is not so high that pulses of adjacent roundtrips overlap. This condition also limits the realized lattice size in our case to about 100 x 100 sites, where limits are given by $x_{\max} \approx \overline{T}/\Delta T_{\text{inner}}$ and $y_{\max} \approx \Delta T_{\text{inner}}/\Delta T_{\text{outer}}$. Additionally, the smallest temporal delay (in this case $\Delta T_{\text{outer}}$) of 30 ns has still to be longer than the pulse duration, which is 22 ns. In our experiment, pulse energies are measured with a photodetector, the signal of which is sampled electronically by an offline digital signal processing software. To reduce noise and to improve the measurement quality of the photodetected pulse sequence, we always average over 100 subsequent single runs of the experiment, each requiring about 30 ms. The final result is mapped according to Eq. (1) onto a two-dimensional discrete lattice spanned by the $x$ and $y$ coordinates (see Fig. 1b,c).

Following the 2D mapping technique, displayed in Fig. 1c, each time a pulse propagates through the shorter instead of the longer inner loop, it advances in arrival time by $\Delta T_{\text{inner}} = 3$ μs or the integer $x$ is decreased by one and vice versa. In the same way, a passage through the shorter (longer) outer loop causes an advance (delay) in arrival time by $\Delta T_{\text{outer}} = 30$ ns, corresponding to a decrease (increases) of $y$ by one. This synthetic lattice displayed in Fig. 1b, is for instance equivalent to a 2D waveguide array and both 50/50 couplers resemble the effect of wave coupling to neighboring waveguides in horizontal or vertical direction. Consequently, each round trip $m$ in our systems can be considered as a propagation by one coupling length in a 2D waveguide array.

In our experiment, pulse dispersion is negligible and each optical pulse in the experiment is well expressed by a single complex amplitude in the simulation and their dynamics are described by four discrete evolution equations, representing pulses travelling to the left ($a_{x,y}^m$ - long inner loop), right ($b_{x,y}^m$ - short inner loop), down ($c_{x,y}^m$ - long outer loop) and upwards ($d_{x,y}^m$ - short outer loop) in an effective square lattice with the transverse coordinates $x$ and $y$. Accordingly, the pulse evolution during the $m^{\text{th}}$ roundtrip is described by a simple set of equations denoting the passage through the inner set of loops as

$$a_{x,y}^m = \tfrac{1}{\sqrt{2}}\left(c_{x-1,y}^{m-1} + id_{x-1,y}^{m-1}\right)\exp\left(i\chi\left|c_{x-1,y}^{m-1} + id_{x-1,y}^{m-1}\right|^2\right) \text{ and} \tag{2}$$

$$b_{x,y}^m = \tfrac{1}{\sqrt{2}}\left(d_{x+1,y}^{m-1} + ic_{x+1,y}^{m-1}\right)\exp\left(i\chi\left|d_{x+1,y}^{m-1} + ic_{x+1,y}^{m-1}\right|^2 + i(-1)^m\varphi_0\right). \tag{3}$$

Followed by the pulse dynamics in the two outer loops as

$$c_{x,y}^m = \tfrac{1}{\sqrt{2}}\left(a_{x,y-1}^m + ib_{x,y-1}^m\right)\exp\left(i\chi\left|a_{x,y-1}^m + ib_{x,y-1}^m\right|^2\right) \text{ and} \qquad (4)$$

$$d_{x,y}^m = \tfrac{1}{\sqrt{2}}\left(b_{x,y+1}^m + ia_{x,y+1}^m\right)\exp\left(i\chi\left|b_{x,y+1}^m + ia_{x,y+1}^m\right|^2 + i(-1)^m\varphi_0\right), \qquad (5)$$

where the first part on the right hand side of Eqs. (2)-(5) represents the interference of pulses from adjacent lattice sites inside the 50/50 coupler. The second one denotes an additional power-dependent nonlinear phase shift ($\varphi_{NL} = \chi|a|^2$) proportional to the effective fiber nonlinearity $\chi$ (see Supplementary Note 3). Phase modulators (PM) placed in the inner and outer short loops induce a phase $(-1)^m\varphi_0$ alternating between subsequent round trips. This causes a jump in phase between neighboring lattice positions acting like an oscillating potential, which opens a tunable gap in the band structure of the system. In addition, we employ erbium-based fiber amplifiers (EDFAs) in each loop to precisely compensate all signal losses caused by absorption and monitoring, thus restoring the conservative nature of the system and enabling a considerable increase of propagation steps[26,27,28], while still being able to follow the whole evolution.

In accordance with a 2D quantum walk dynamics[24,25], the first experiment ($\varphi_0 = 0$) shows a two-dimensional light walk at low initial input power ($\varphi_{NL} \approx 0$) in Fig. 2. An initial single pulse (see Fig. 2a) injected into the left outer loop (on the center of the lattice $x = 0, y = 0$, see Fig. 2c) results in a one-dimensional stream of arrival pulses (see Fig. 2b), which are sampled and attributed to the correspondent 2D synthetic lattice positions (see Fig. 2c-f). Note that the expanded inset of Fig. 2a,b (dotted red square) represents all values displayed on the $x$-axis for $y = 0$. This successful emulation of further synthetic dimensions by using a genuine 1D fiber is based on the combination of long and short range coupling caused by the inner and outer loop pairs, respectively.

Next, we demonstrate that spatial soliton formation and nonlinearly induced beam collapse, as it happens in bulk materials, can also be realized in an artificial 2D synthetic system based on coupled fiber loops. In the linear case ($\chi = 0$), our discrete system has a band structure and its eigenmodes are represented by a Floquet-Bloch ansatz

$$\begin{pmatrix} a_{x,y}^m \\ b_{x,y}^m \end{pmatrix} = \begin{pmatrix} A \\ B \end{pmatrix} e^{i/2(-\theta m + k_x x + k_y y)}. \qquad (6)$$

which encodes the phase acquired by a Bloch wave in a double step of Eqs. (2)-(5). Here, we take into account that a pulse requires two round trips in order to return to an initial point (for instance left, up, right and downwards) and that our unit cell is therefore two elements wide. The Floquet phase $\theta$ and the Bloch momenta $k_x$ and $k_y$ obey the dispersion relation

$$\cos\theta = \tfrac{1}{2}\left[-\tfrac{1}{2} - \tfrac{1}{2}\cos(2\varphi_0) - \cos\varphi_0 \cos k_x - \cos\varphi_0 \cos k_y + \cos k_x \cos k_y\right], \qquad (7)$$

which can be easily tuned by the alternating the phase modulation $\varphi_0$. For $\varphi_0$ different from 0 or $\pi$, two mirror symmetric bands separated by a gap appear (see Fig. 3b).

As solitons are localized states with exponentially decaying tails, they can only exist in the presence of a band gap, in which their propagation constant is situated. As nonlinearity shifts propagation constants away from linear waves in a defined direction, a certain sign of the band curvature is mandatory for the soliton's propagation constant to reach the gap. As shown in Fig. 3b, the two bands of our system feature opposite curvatures. As the fiber nonlinearity is positive, only the upper band allows for soliton formation when excited in the center of the Brillouin zone by a broad Gaussian wave packet with narrow momentum spread (see Fig. 3b). Here, we choose an overall phase modulation of $\varphi_0 = \pi/2$, which guarantees a broad gap of width $\Delta\theta = 2\pi/3$ and for which the upper band has a constant positive curvature in a wide momentum range and thus also resembles the dispersion relation of waves propagating in bulk materials (see Fig. 3b).

To investigate the wave propagation, we launch a sequence of pulses corresponding to a Gaussian envelope $G_w(x, y) = A_w \exp[-(x^2 + y^2)/w^2]$ along $x$ and $y$ (see Fig. 4a-c) with a variable amplitude $(A_w)$ and a fixed width $(w)$ of 6 positions ($1/e$ drop of intensity) and realize a selective excitation centered on the $\Gamma$ point $(k_x = 0, k_y = 0)$ of the upper band (for more details of the preparation, see Supplementary Note 5). At low optical power (0.3 mW), the field distribution spreads very similar to a Gaussian beam diffracting in free space (see Figs. 4d-f) to a width of about 9 positions. But, originally this spreading is based on pulses acquiring new positions in the time domain (compare Fig. 4d with Fig. 4a).

For a medium power level (1 mW), the positive nonlinearity of the fiber shifts nonlinear waves into the gap, thus isolating them from the linear spectrum and providing the conditions for localization. Hence, nonlinearity starts suppressing the coupling (see Fig. 4g), resulting in the formation of a solitonic structure (see Figs. 4h,i), which propagates for up to 60 time steps in the experiment, provided that the initial power has been adjusted carefully. Yet, even in computer simulations, numerically optimized soliton profiles finally decompose and either self-compress or broaden. Note that this behavior is well known for the unstable Townes soliton of the continuous 2D NLS[29,30] (see Supplementary Note 6).

To illuminate the soliton behavior, we have determined numerically the whole family of lattice solitons inside the band gap (see fig. 5k). For propagation constants close to the upper band $\theta \rightarrow \theta_{\text{linear}} = \pi/3$, we find solitons with infinite width and a shape approaching that of their continuous counterpart the Townes-like soliton. As expected, these solutions are weakly unstable and this instability is even boosted by discreteness. However, resulting life times are still much larger than those accessible in the experiment, which explains the experimental accessibility. Only for soliton propagation constants within the lower half of the gap, any perturbation results in an almost immediate blow up of the field, causing a transformation into a less energetic and more stable highly localized stationary solution (see Supplementary Note 6).

By choosing even higher input power (3 mW, see Figs. 4j-l), nonlinear self-focusing dominates already at the beginning, thus resulting in a strong contraction of the field distribution (compare linear and nonlinear propagation plots in Fig. 5a-j). As our system is inherently discrete, this collapse is stopped at the single lattice site. While some excess radiation is shed away into the lattice, most of the power remains concentrated, thus forming an extremely localized and almost stationary state. Interesting to note that in our system such self-compressing instability corresponds to a transfer of all the energy of a sequence of pulses into a single high-power flash of light (see Fig. 4j).

This extreme self-localization is generic to quasi-two-dimensional nonlinear systems with a gap and does not depend on the shape of the initial excitation. If we insert a single pulse into the left outer loop (see Fig. 6a), corresponding to the center of the lattice ($x=0, y=0$, see Fig. 6b,c), we excite the whole band structure. As expected, a low power (0.3 mW) excitation spreads on the 2D lattice (see Fig 6d-f). But, above a certain power threshold, diffraction is compensated and the same extremely localized state appears (see Fig. 6g-i).

In conclusion, we successfully implemented two transverse synthetic dimensions in a fiber optical system by combining short and long-range coupling, thus resulting in a 100 x 100 lattice. We realized an eigenvalue spectrum containing two bands separated by a gap and thus induced dispersion, which is partially similar to that of bulk materials. Consequently, we observed the collapse of a broad initial field distribution around a single lattice site caused by the action of a cubic nonlinearity, which corresponds to a pulse compression scheme working at mW-power levels. As our set-up is extremely flexible, it can be easily extended towards other challenging lattice models, such as 2D topological insulators or non-conservative system with even more synthetic dimensions.

## Acknowledgements

This project was supported by the German Research Foundation (DFG) through the International Research Training Group (IRTG) 2101, as well as an NSERC CREATE grant. RM also acknowledges additional support by the Government of the Russian Federation through the ITMO Fellowship and Professorship Program (grant 074-U 01) and by the 1000 Talents Sichuan Program in China.

## Author contributions

A. L. M. M. performed the measurements, M.W. designed the experimental setup, A.B. performed simulations. R.M. and U.P. developed the theoretical background.

## Additional Information

**Supplementary information** accompanies this paper at …

**Competing Interests:** The authors declare no competing interests.

**Publisher's note:** Springer Nature remains neutral with regard to jurisdictional claims in published maps and institutional affiliations.

**Figures**

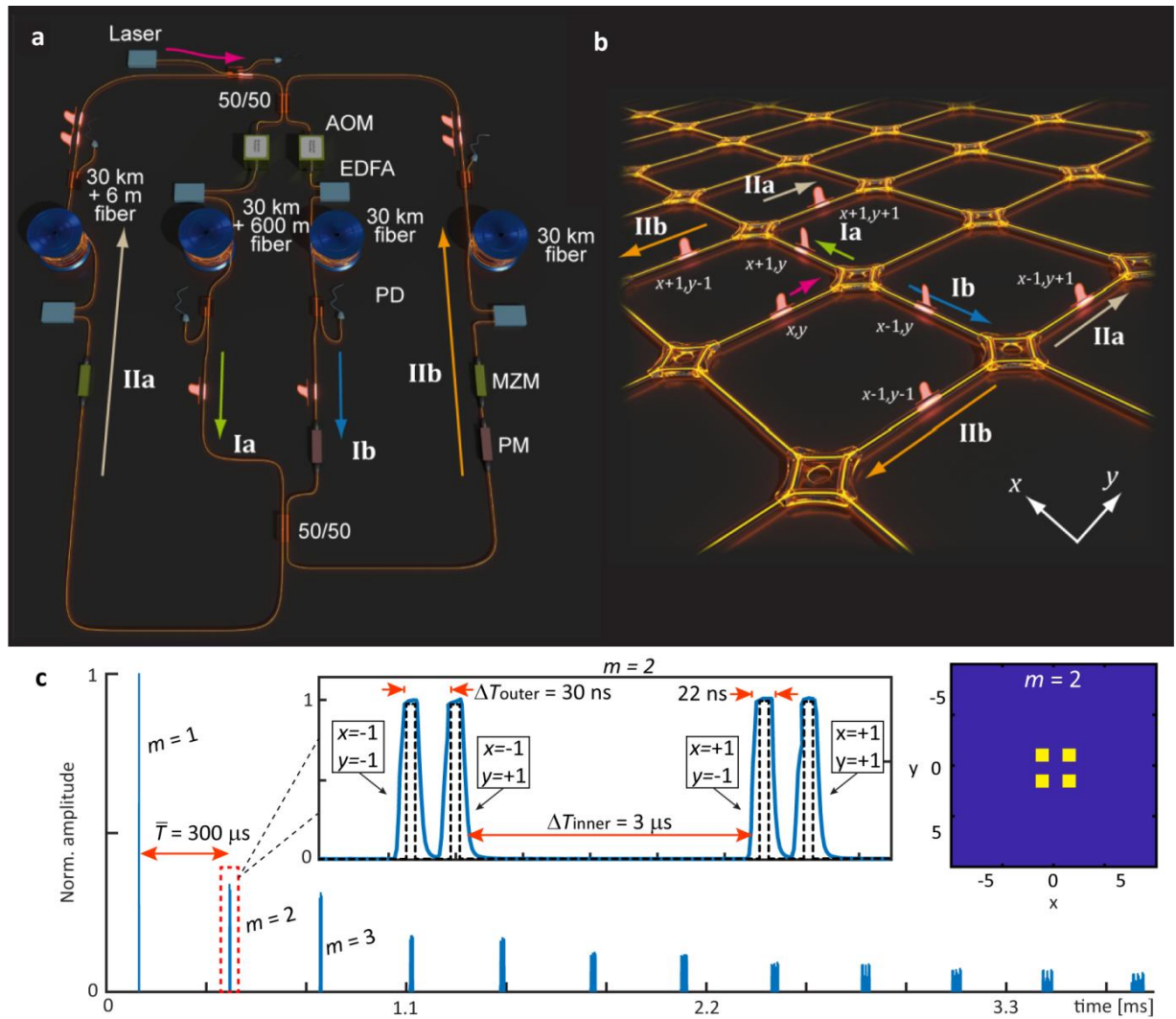

**Figure 1. Experimental realization of light propagation on a 2D mesh lattice. a**) Experimental set-up consisting of two pairs of fibers connected via 50/50 couplers. In front of each loop, consisting of about 30 km of optical fiber, there is an erbium-doped fiber amplifier (EDFA) for loss compensation. Acoustic optical modulators (AOMs) in the inner and Mach-Zehnder modulators (MZMs) in the outer pair allow for amplitude modulation. Additionally, phase modulators (PMs) are placed in the inner and outer right loop. **b**) and **c**) Equivalent 2D lattice and measured time trace at the photodetector (PD) of the inner loop: A pulse is created and injected into the outer left loop. During each round trip, it splits in the fiber couplers and generates four smaller pulses with different arrival times according to the fiber pieces they have passed. The power of the pulses is measured in each round trip $m$ (blue line), sampled electronically (black dashed line) by a computer software and mapped onto a 2D lattice according to the arrival times (here displayed for $m$=2) Thus pulses effectively propagate on a 2D synthetic lattice where a step in + / - $x$ direction corresponds to a passage through the longer /shorter inner loop and transport in + / - $y$ – direction is realized by transits through the long / short outer loop.

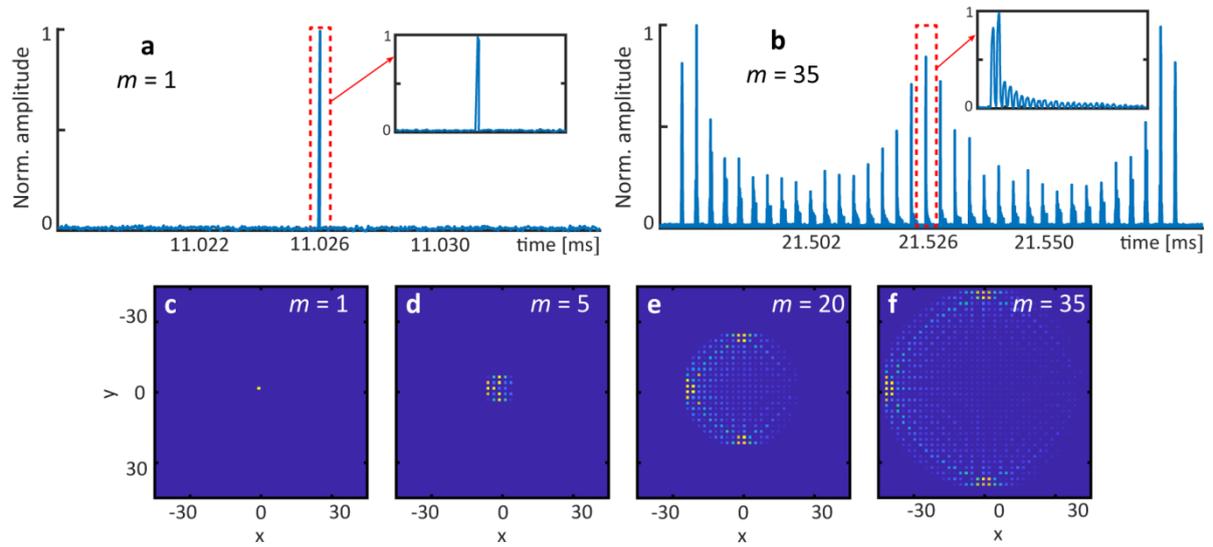

**Figure 2. Experimental realization of a 2D Light walk for vanishing phase modulation $\varphi_0$ = 0. a)** Photodetected signal of the first time step $m$, at which a single pulse is inserted into the left outer loop. **b)** Oscilloscope trace of time step $m = 35$. **c-f**, 2D Light walk for different propagation steps $m$.

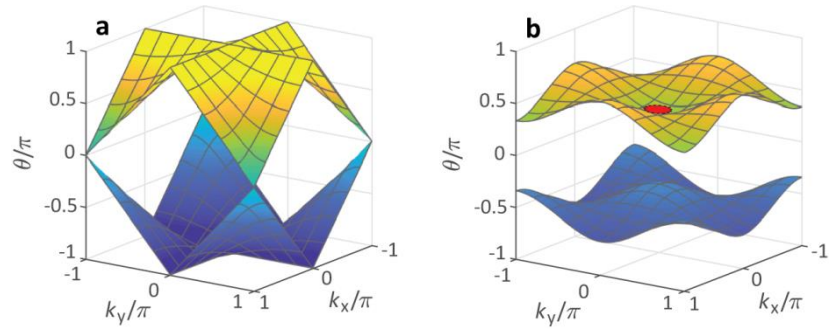

**Figure 3. Band structure of the 2D mesh lattice.** Due to the Floquet nature of the system, not only the Bloch momenta $k_{x/y}$, but also the propagation constant $\theta$ are periodic within $[-\pi; \pi]$. **a)** Passive band structure in absence of any phase modulation. **b)** By applying a phase modulation $\varphi_0 = \pi/2$ alternating every time step, the Dirac cone at $k_x = k_y = \pm\pi$ opens up and a gap appears. The red circle on the upper band represents a selective excitation of a Gaussian wave packet in the center of the Brillouin Zone.

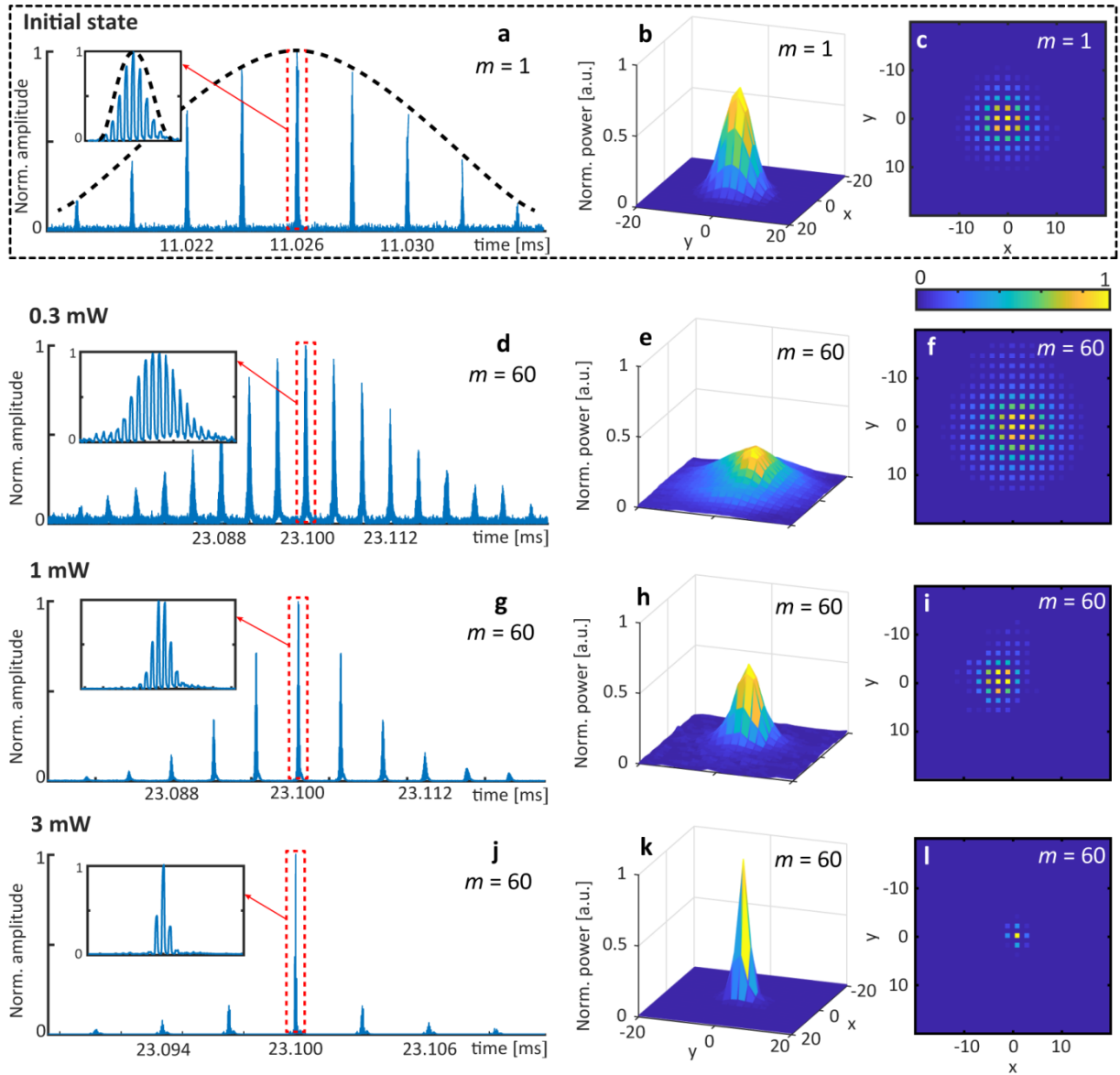

**Figure 4. Experimentally detected evolution of a broad excitation in the presence of a gap ($\varphi_0 = \pi/2$) and for different power levels demonstrating wave collapse at mW power level (bottom row).** The system is excited with a sequence of pulses with a Gaussian envelope (black dashed line) as shown in the (a) time domain, (b) a 3D surface plot and (c) a 2D image with normalized scaled colors. **d-f** Linear diffraction of a Gaussian beam on the lattice (time step $m = 60$, 0.3 mW input power). **g-i** Townes-like soliton for intermediate power (time step $m = 60$, 1.0 mW input power). **j-l** Collapse of the field distribution into a single lattice site at the highest power level (time step $m = 60$, 3.0 mW input power).

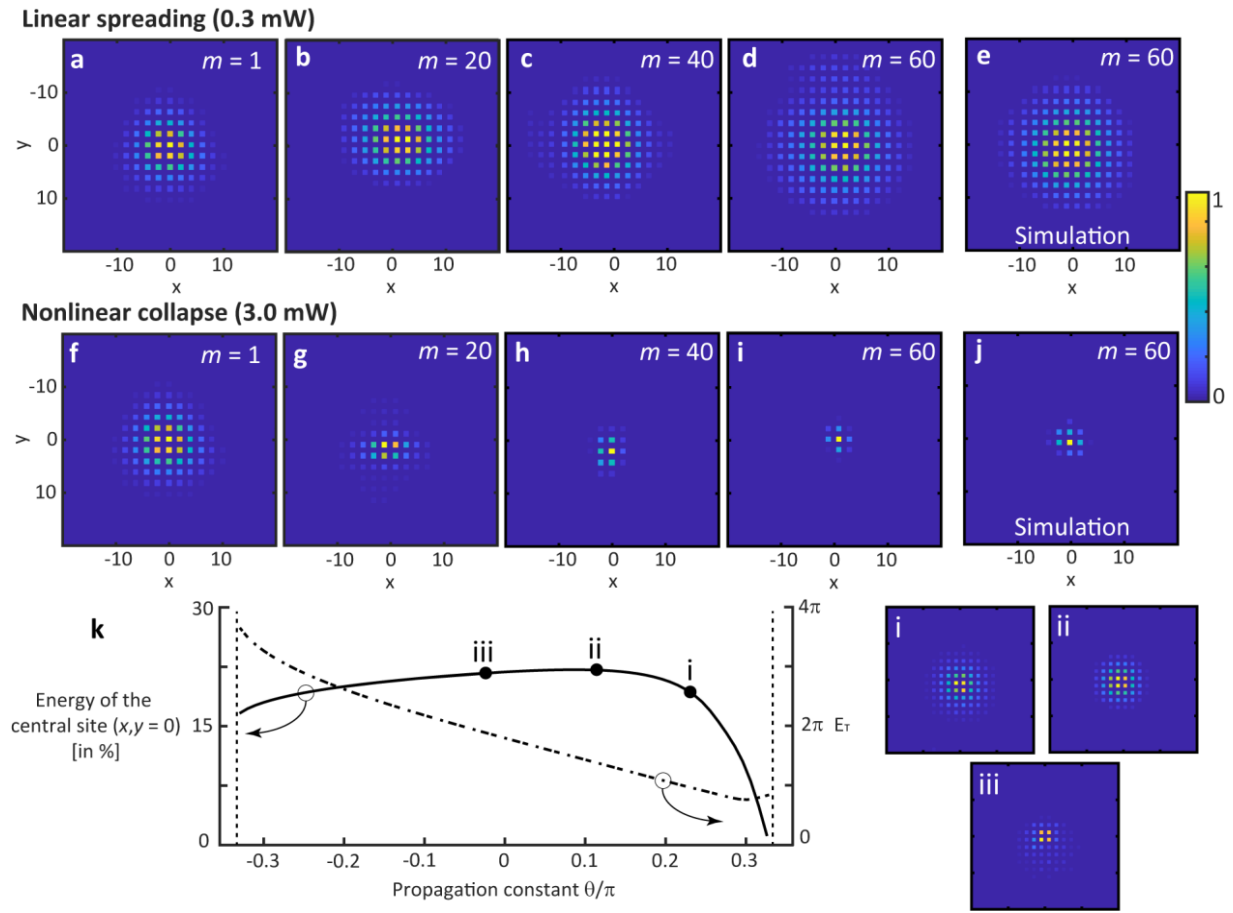

**Figure 5.** Propagation plots of linear (input power $0.3$ mW) spreading (a-d) and nonlinear (input power $3$ mW) collapse (f-i) of a broad excitation ($\varphi_0 = \pi/2$) in agreement with numerical simulations (e,j). **k)** Fraction of total energy situated in the central side and total energy for a numerically determined soliton profile as a function of propagation constant (nonlinear coefficients $\chi$ was set to unity, respective soliton profiles are given on the right).

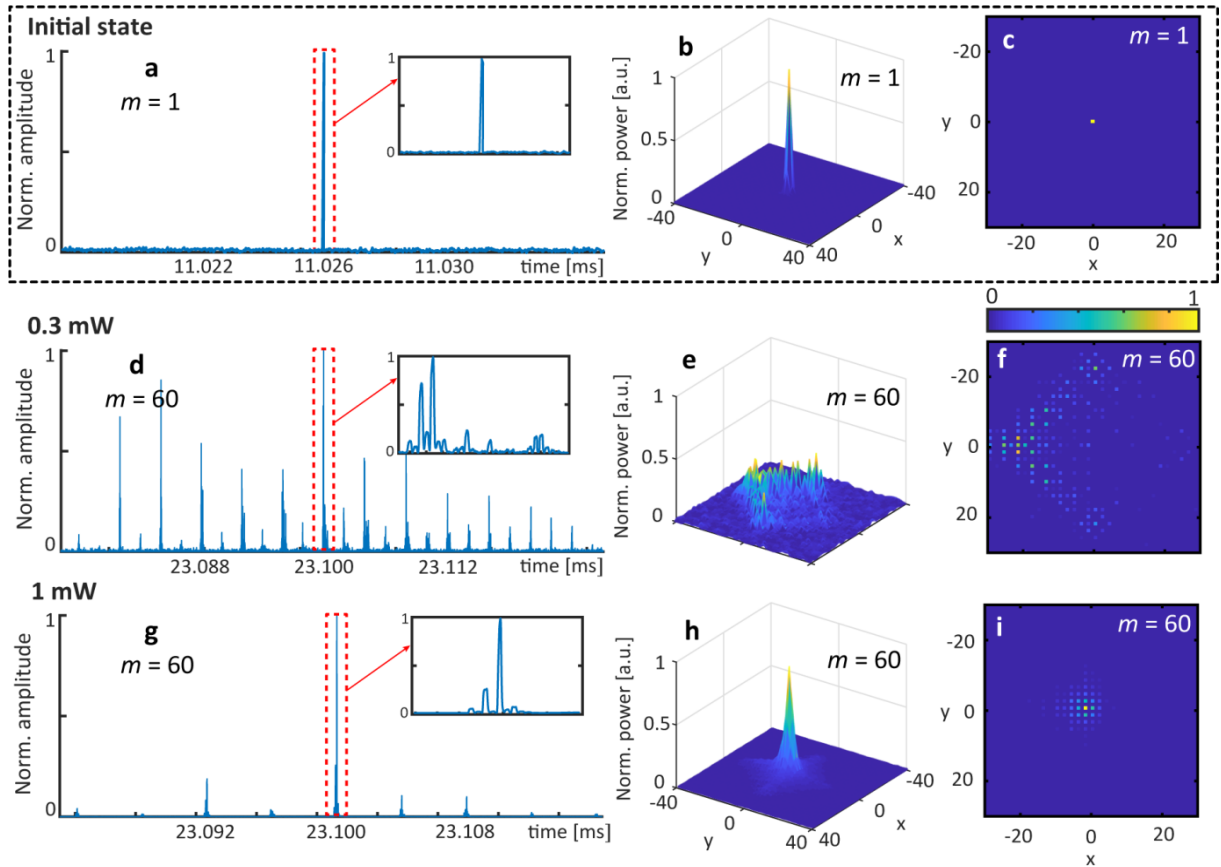

**Figure 6. Evolution of a single lattice excitation in the presence of a gap ($\varphi_0 = \pi/2$) demonstrating nonlinearly induced localization for high input power (bottom row).** The system is excited with a single pulse as shown in the (a) time domain, (b) a 3D surface plot and (c) a 2D image with normalized scaled colors. **d-f,** Linear pulse spreading onto the lattice (time step $m = 60$, 0.3 mW input power). **g-i,** Nonlinearly driven localization at a single lattice site (time step $m = 60$, 1 mW input power).